\documentclass[final,5p,times,twocolumn]{elsarticle}

\usepackage{graphicx}
\graphicspath{{./graphics/}}
\usepackage{amsmath}
\usepackage{textcomp}

\ifcsname italic \endcsname
\else
  
\fi

\journal{Nuclear Instruments and Methods A}

\begin{document}
\begin{frontmatter}

\title{A novel technique for determining luminosity in electron-scattering/positron-scattering experiments
from multi-interaction events}
\author[mit]{A.~Schmidt\corref{cor}}
\ead{schmidta@mit.edu}

\author[mit]{C.~O'Connor}
\author[mit]{J.\ C.~Bernauer}
\author[mit]{R.\ Milner}

\address[mit]{Massachusetts Institute of Technology, Laboratory for Nuclear Science, Cambridge, MA, USA}

\cortext[cor]{Corresponding author}

\begin{abstract}
  The OLYMPUS experiment measured the cross-section ratio of positron-proton elastic
  scattering relative to electron-proton elastic scattering to look for evidence of
  hard two-photon exchange. To make this measurement, the experiment alternated between
  electron beam and positron beam running modes, with the relative integrated luminosities of
  the two running modes providing the crucial normalization. For this reason, 
  OLYMPUS had several redundant luminosity monitoring systems, including a pair
  of electromagnetic calorimeters positioned downstream from the target to detect
  symmetric M\o ller and Bhabha scattering from atomic electrons in the hydrogen
  gas target. Though this system was designed to monitor the rate of events with 
  single M\o ller/Bhabha interactions, we found that a more accurate determination
  of relative luminosity could be made by additionally considering the rate of 
  events with both a M\o ller/Bhabha interaction and a concurrent elastic $ep$
  interaction. This method was improved by small corrections for the variance
  of the current within bunches in the storage ring and for the probability of three
  interactions occurring within a bunch. After accounting for systematic effects, 
  we estimate that the method is accurate in determining the relative luminosity 
  to within 0.36\%. This precise technique can be employed in future electron-proton
  and positron-proton scattering experiments to monitor relative luminosity between different
  running modes.
\end{abstract}


\end{frontmatter}

\section{Introduction}

The OLYMPUS experiment \cite{Milner:2013daa} measured the ratio of positron-proton to electron-proton 
elastic scattering cross-sections for a 2.01~GeV beam energy over a range of scattering angles \cite{Henderson:2016dea}.
The analysis depended on the accurate (better than $1\%$) determination of the relative integrated luminosity between 
positron-beam running mode and electron-beam running mode. Three luminosity monitoring systems
were implemented for the experiment, of which one was a pair of Symmetric M\o ller/Bhabha (SYMB) calorimeters,
designed to use the rate of symmetric scattering from the atomic electrons in the hydrogen
target to extract luminosity \cite{Benito:2016cmp}. Since the M\o ller and Bhabha scattering cross-sections can be calculated 
precisely using quantum electrodynamics, this method was expected to have high performance. M\o ller and 
Bhabha calorimeters had previously been used successfully for luminosity monitoring with the HERMES experiment
\cite{Benisch:2001rr}.

Unfortunately, this method turned out to be ill-suited for the needs of OLYMPUS. Whereas HERMES typically 
needed the relative luminosity between running modes with two different beam or target polarization states
using the same beam species, OLYMPUS required the relative luminosity between running modes with different beam species. 
Since the M\o ller and Bhabha differential cross-sections have different angular dependences close to the symmetric
angle, this method required control of the calorimeter acceptance to a better degree than was possible, even
with sophisticated simulations. We estimated the accuracy of the relative luminosity determination from SYMB 
to be on the order of 4\%. 

Nevertheless, we developed an alternate method to determine the relative luminosity from the SYMB data, 
achieving an accuracy of 0.36\% by comparing the rate of symmetric M\o ller/Bhabha events to the rate of 
a specific type of multi-interaction event (MIE). In this specific type of MIE, a symmetric M\o ller/Bhabha 
interaction occurs simultaneous to an elastic lepton-proton scattering event in which the lepton enters one 
of the two calorimeters. While the rate of M\o ller/Bhabha events scales with luminosity, the MIE rate scales 
with luminosity squared, and by taking a ratio, the luminosity can be recovered. This MIE method has three 
principal advantages that make it robust against many systematic effects:
\begin{enumerate}
\item The important quantity is a ratio of rates rather than a single rate, canceling some systematics,
\item The ratio is nearly the same in both electron and positron modes,
\item There is a reduced burden on acceptance simulations.
\end{enumerate}

In this paper, we present a derivation of the MIE method, estimate its associated systematic accuracy for OLYMPUS, 
and discuss how it might be useful for luminosity monitoring in future experiments. 

\section{The OLYMPUS Experiment}

\begin{figure*}[t]
\centering
\includegraphics[width=0.9\textwidth]{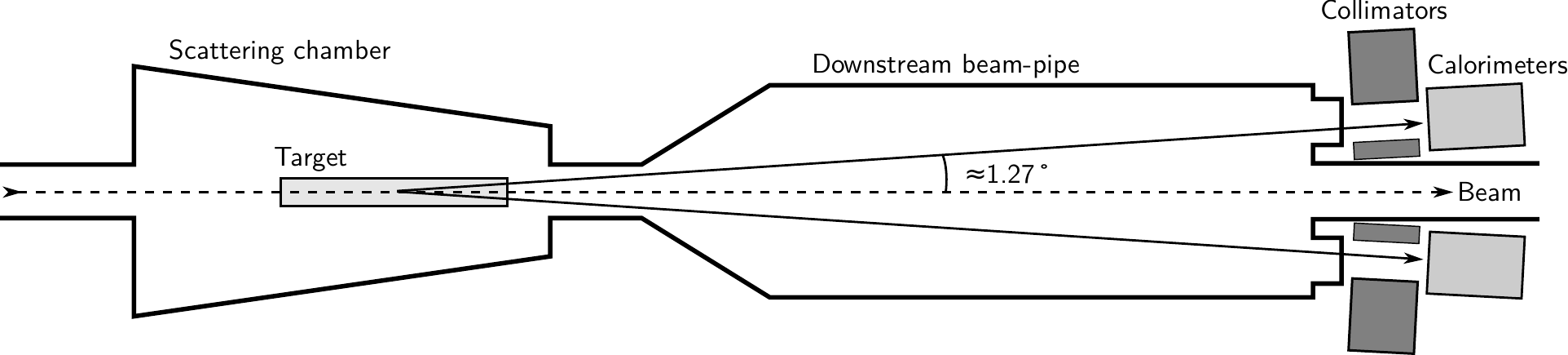}
\caption{\label{fig:schematic} The SYMB calorimeters were positioned approximately 3~m downstream from
the target, on either side of the beamline, as seen in this schematic (not to scale).}
\end{figure*}

The OLYMPUS experiment collected data at the DORIS storage ring, at DESY, Hamburg, in 2012. 
DORIS was capable of storing both electron and positron beams, and, in OLYMPUS, the beam species
was switched approximately once per day. Determining the relative integrated luminosity of the data 
from the two different beam species was crucial for the OLYMPUS measurement. 

OLYMPUS operated with a fixed hydrogen gas target. The SYMBs were positioned approximately 3~m downstream 
from the target, at an approximate scattering angle of $1.27^\circ$. This corresponded to the symmetric angle
for M\o ller and Bhabha scattering at the beam energy of 2.01~GeV. The two calorimeters were placed on either side of the beam line in order
to detect both final state leptons in coincidence. Fig.~\ref{fig:schematic} shows a schematic of calorimeter
placement relative to the target and beamline. Relevant distances and angles, determined from the OLYMPUS
optical survey, are given in Table \ref{tab:symb}.

\begin{table}[htpb]
\caption{ \label{tab:symb} Relevant geometric parameters for the SYMB calorimeters }
\centering
\begin{tabular}{| l l |}
\hline
Parameter & Value \\
\hline
\hline
Dist. to upstream face of collimator & 2992~mm\\
Dist. to upstream face of calorimeter & 3117~mm\\
Collimator thickness & 100~mm \\
Radius of collimator aperture & 10.25~mm \\
Angle to center of left collimator aperture & $1.32^\circ$ \\
Angle to center of right collimator aperture & $1.28^\circ$ \\
Orientation of left collimator aperture & $1.30^\circ$\\
Orientation of right collimator aperture & $1.22^\circ$\\
M\o ller-Bhabha symmetric angle & $1.27^\circ$ \\
\hline
\end{tabular}
\end{table}

There were two other luminosity monitoring systems in OLYMPUS in addition to the SYMBs.  The first was a pair of 
forward tracking telescopes, which monitored the rate of forward lepton-proton scattering. This method of determining 
luminosity was relative to the asymmetry, if any, between the electron-proton and positron-proton cross-sections---such 
as that caused by hard two-photon exchange (TPE)---at the forward scattering angle. However, given a luminosity
determination from the SYMBs, the forward tracking telescopes could make a determination of hard TPE. This determination
was reported in the OLYMPUS results \cite{Henderson:2016dea}. The second system was the OLYMPUS slow control system,
which recorded the beam current in the storage ring, the flow rate of hydrogen to the target, and the target temperature,
which could be combined into a luminosity determination. This method was only accurate to the order of $3\%$, but had the
advantage that it could be made online during data taking, without any simulation or track reconstruction.

\begin{figure}[htpb]
\centering
\includegraphics{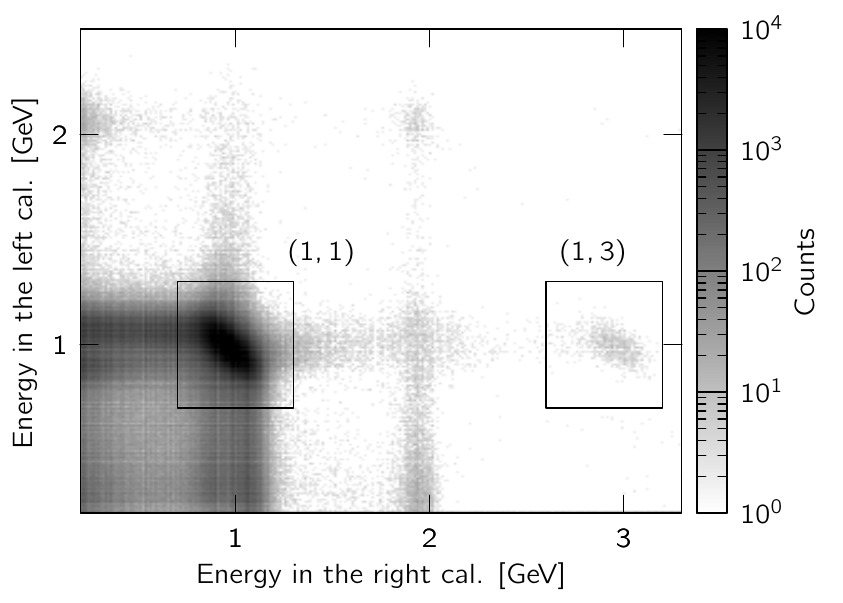}
\caption{\label{fig:lm_hist} Luminosity can be extracted from the ratio of $(1,3)$ MIE events (in the right
box) to single $(1,1)$ M\o ller/Bhabha events (in the left box).}
\end{figure}

The SYMB data were digitized using fast histogramming cards, whose acquisition time was less than the $\approx 100$~ns bunch
separation time at DORIS. The system was dead-time free. The cards provided two-dimensional histograms, in which one axis
corresponded to the energy deposited in the left calorimeter and the other axis corresponded to the energy of the right
calorimeter. Each fill of the histogram corresponded to one beam bunch passing through the target. The SYMB data consisted
of three histograms, one triggered on the left calorimeter, one triggered on the right calorimeter, and one triggered
on left-right coincidences. The three histograms had different dynamic ranges. The dynamic range of the coincidence histogram
was optimized to cover the energy range of symmetric M\o ller/Bhabha events. 
The left- and right-triggered histograms, which were intended as diagnostics, had larger dynamic ranges to cover
forward-going elastic $ep$ events. By chance, the dynamic range of the left-triggered histogram
was large enough to include events of more than 3~GeV, permitting the analysis discussed in this work. 
The left-triggered histgram for a subset of data is shown in Fig.~\ref{fig:lm_hist}. 
There are several dense areas---signal peaks---that correspond to specific processes. The most prominent is due 
to symmetric M\o ller and Bhabha scattering, which deposits approximately 1 GeV in both the left and the right calorimeters.

\section{Derivation}

In this derivation, we consider three types of events of interest. In symmetric M\o ller/Bhabha events, denoted
$ee$, 1~GeV of energy is deposited in each calorimeter. We denote elastic lepton-proton interactions in which
the lepton deposits 2~GeV in the left calorimeter with $ep\rightarrow L$. We denote elastic lepton-proton interactions
in which the lepton deposits 2~GeV in the right calorimeter with $ep\rightarrow R$. Since the energy depositions
in GeV are approximately integer values, we will use a coordinate system $(L,R)$ to label signal peaks. For example,
peak $(1,3)$ refers to the signal peak with 1~GeV deposited in the left and 3~GeV deposited in the right.

Since the number of beam particles in a bunch is large (order $10^{10}$) while the probability of an interaction
occurring in a given bunch is small, the probability of having $N$ interaction in a bunch with integrated luminosity 
$\mathcal{L}_j$ follows a Poisson distribution,
\begin{equation}
P(N) = \frac{e^{-\sigma \mathcal{L}_j} (\sigma \mathcal{L}_j)^N}{N!},
\end{equation}
where $\sigma$ is the cross-section for an interaction. The probability for a bunch crossing to result in an event
of form $(1,1)$ is:
\begin{align}
P(1,1) =& P_{ee}(1) \times P_{ep\rightarrow L}(0) \times P_{ep\rightarrow R}(0) \\
= & e^{-\mathcal{L}_j (\sigma_{ee} + \sigma_{ep\rightarrow L} + \sigma_{ep\rightarrow R})} \times \sigma_{ee} \mathcal{L}_j\\
= & e^{-\mathcal{L}_j \sigma_\text{tot.} } \times \sigma_{ee} \mathcal{L}_j,
\end{align}
where we define $\sigma_\text{tot.} \equiv \sigma_{ee} + \sigma_{ep\rightarrow L} + \sigma_{ep\rightarrow R}$. 
In the same way, we can write the probability for energy deposition of the form $(1,3)$:
\begin{align}
P(1,3) =& P_{ee}(1) \times P_{ep\rightarrow L}(0) \times P_{ep\rightarrow R}(1) \\
= & e^{-\mathcal{L}_j (\sigma_{ee} + \sigma_{ep\rightarrow L} + \sigma_{ep\rightarrow R})} \times \sigma_{ee} \sigma_{ep\rightarrow R} \mathcal{L}_j^2 \\
= & e^{-\mathcal{L}_j \sigma_\text{tot.} } \times \sigma_{ee} \sigma_{ep\rightarrow R} \mathcal{L}_j^2.
\end{align}

Over a run period with $N_b$ bunches crossing the target, the expected number of events, $N$, in a given signal peak can
be found by summing the probabilities over each bunch (assuming $N_b$ is sufficiently large):
\begin{align}
N_{(1,1)} =& \sum\limits_{j=0}^{N_b} P(1,1) =  \sum\limits_{j=0}^{N_b} \left[
  e^{-\mathcal{L}_j\sigma_\text{tot.} } \times \sigma_{ee} \mathcal{L}_j \right], \\
N_{(1,3)} =& \sum\limits_{j=0}^{N_b} P(1,3) =  \sum\limits_{j=0}^{N_b} \left[
  e^{-\mathcal{L}_j \sigma_\text{tot.}} \times
  \sigma_{ee} \sigma_{ep\rightarrow R} \mathcal{L}_j^2 \right].
\end{align}
Since $\mathcal{L}_j \sigma_\text{tot.} \ll 1$, we can recast the exponentials as a power series, which we can truncate.
Dividing the two rates, dropping terms beyond first order in $\mathcal{L}_j \sigma_\text{tot.}$, and using 
$\mathcal{L}=\sum_j^{N_b}\mathcal{L}_j$ to represent the integrated luminosity of the entire run period, we get:
\begin{align}
\frac{N_{(1,3)}}{N_{(1,1)}} &= \frac{ \sigma_{ep\rightarrow R} \sum\limits_{j=0}^{N_b} \mathcal{L}_j^2
\left[ 1 - \mathcal{L}_j\sigma_\text{tot.} \right]}
{\sum\limits_{j=0}^{N_b} \mathcal{L}_j \left[ 1 - \mathcal{L}_j\sigma_\text{tot.} \right]}, \\
&= \frac{ \sigma_{ep\rightarrow R} N_b \left[ \langle \mathcal{L}_j^2 \rangle - \sigma_\text{tot.} \langle \mathcal{L}_j^3 \rangle \right] }
{\left[ \mathcal{L} - N_b \sigma_\text{tot.} \langle \mathcal{L}_j^2 \rangle \right]},\\
&= \frac{ \sigma_{ep\rightarrow R} N_b }{\mathcal{L}}
\left[ \langle \mathcal{L}_j^2 \rangle - \sigma_\text{tot.} \langle \mathcal{L}_j^3 \rangle \middle]
\middle[ 1 + \frac{N_b}{\mathcal{L}} \sigma_\text{tot.} \langle \mathcal{L}_j^2 \rangle \right],
\end{align}

The variance in luminosity per bunch, $v_b$, is equal to $\langle \mathcal{L}_j^2 \rangle - \langle \mathcal{L}_j \rangle ^2$, or equivalently
$\langle \mathcal{L}_j^2 \rangle - \mathcal{L}^2/N_b^2$. Using this substitution, dropping the $\sigma_\text{tot.}^2$ term, and rearranging
we arrive at:
\begin{equation}
\mathcal{L} = \frac{N_{(1,3)}N_b }{N_{(1,1)} \sigma_{ep\rightarrow R} }
- \frac{v_b N_b^2}{\mathcal{L}} - N_b \sigma_\text{tot.}
  \left\{ \left(\frac{v_b N_b}{\mathcal{L}} + \frac{\mathcal{L}}{N_b}\right)^2 - \frac{\langle \mathcal{L}_j^3 \rangle N_b}{\mathcal{L}}\right\}.
\label{eq:mie_result}
\end{equation}
Eq.~\ref{eq:mie_result} shows that luminosity can be estimated from a main
term, $N_{(1,3)} N_b / N_{(1,1)} \sigma_{ep\rightarrow R}$, with some corrections.
In this derivation, only terms to first order in $\mathcal{L} \sigma_\text{tot.}$,
have been kept, but, if necessary, corrections can be calculated to arbitrary order.

Let us consider the meaning of the correction terms. The first correction term, $v_b N_b^2 / \mathcal{L}$,
describes the effect of luminosity variance between bunches. High-luminosity bunches are
more likely to have multi-interaction events than low-luminosity bunches. If there is variance
between the bunches, this will have an effect on the luminosity determination. The second correction
term, which has a leading factor of $\sigma_\text{tot.}$, accounts for the fact that there
may be three interactions in a bunch crossing. Some small fraction of would-be (1,3) events
will fall outside of the (1,3) peak because of additional energy deposited by a third
interaction in the same bunch. 

\begin{figure}[htpb]
\centering
\includegraphics{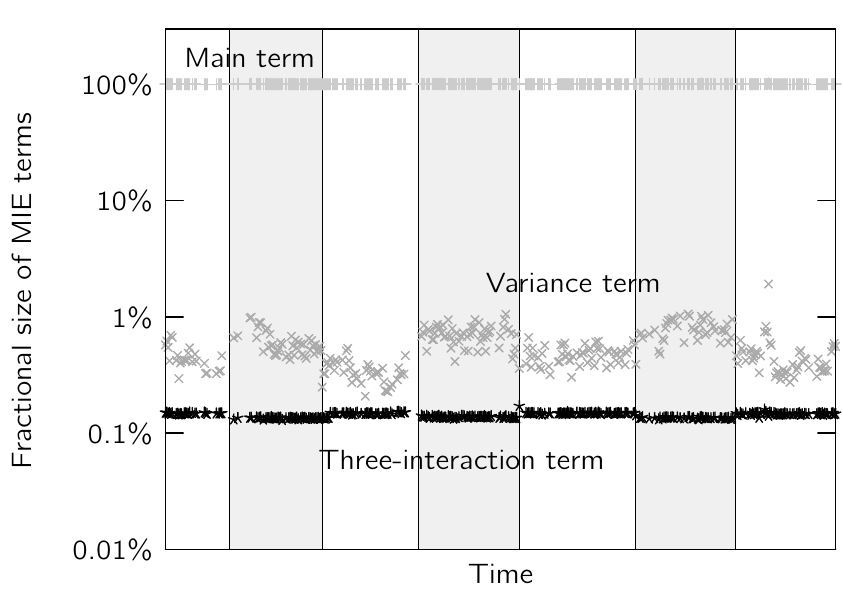}
\caption{\label{fig:mie_terms} The variance term modified the luminosity by slightly less than
a percent, while the three-interaction term was closer to a 0.1\% correction. The background 
shading shows the different beam species; a white background indicates electron beam running, while
a light gray background indicates positron beam running. } 
\end{figure}

Fig.~\ref{fig:mie_terms} shows the scale for these correction terms relative to the total luminosity,
for a typical week of running.
The variance term is highly species-dependent, and corrects the extracted luminosity by a few percent. 
This term is absolutely necessary to achieve percent-level accuracy desired for OLYMPUS. The correction for 
three-interactions per bunch is smaller, on the order of one to two tenths of a percent. Subsequent higher-order
terms are even smaller and can be safely neglected. 

Eq.~\ref{eq:mie_result} is not the result of solving for $\mathcal{L}$; luminosity appears in both
sides of the equation. However, $\mathcal{L}$, $v_b$, and $\langle \mathcal{L}_j^3 \rangle$ appear only in
the correction terms. Since these terms are already small, there is little residual error in using a rough 
estimate of the luminosity to calculate these terms. For the purposes of OLYMPUS, the slow control system 
was suitable for estimating $\mathcal{L}$, $v_b$, and $\langle \mathcal{L}_j^3 \rangle$, as well as $N_b$ 
that appear in the right side of the equation. The cross-sections $\sigma_{ep\rightarrow L}$, 
$\sigma_{ep\rightarrow R}$, and $\sigma_{ee}$ were estimated using simulations. The yields $N_{(1,1)}$ and 
$N_{(1,3)}$ were, naturally, taken from the calorimeter data.

The luminosity determination in this work relies on the $(1,3)$ signal peak, visible only in the left-triggered
histogram, but it would be equally possible to extract luminosity from counts in the $(3,1)$ peak. This was not 
possible for OLYMPUS because the $(3,1)$ signal peak was outside of the dynamic ranges of all three SYMB data histograms. 
If a luminosity monitor were designed around this principle, a comparison of the $(1,3)$ and $(3,1)$ rates would provide
a valuable cross check.

\section{Systematics}

\begin{table}
\centering
\caption{\label{table:mie_sys} Contributions to the systematic uncertainty}
\begin{tabular}{ | l | l | }
\hline
Uncertainty & Value (\%) \\
\hline
\hline
Statisics & 0.10 \\
\hline
Beam Position Monitors & 0.21 \\
Magnetic Field at Target & 0.20 \\
Collimator Geometry & 0.13 \\
Event Selection & 0.10 \\
Downstream Magnetic Field & 0.05 \\
Higher Order Corrections & 0.05 \\
Radiative Corrections & 0.03 \\
\hline
\hline
Total & 0.36 \\
\hline
\hline
Beam Energy & 0.10 \\
\hline
Total incl. Beam Energy & 0.37 \\
\hline
\end{tabular}
\end{table}

In this section, we present our estimate for the systematic uncertainty of the MIE method as applied in
OLYMPUS. The principal source of uncertainty was the use of a simulation to estimate $\sigma_{ep\rightarrow R}$.
Any inaccuracy in the simulation contributed to error in the relative luminosity determination. This
uncertainty estimation is specific to the OLYMPUS simulation as implemented in the OLYMPUS analysis. 
If the MIE method were used in a future experiment, it may prove more or less accurate depending on that
experiment's simulation.

The sources of systematic uncertainty in the simulation can be divided into several classes, which we will address in turn. 
Our estimates of the uncertainty contributed by each of these classes is shown in Table
\ref{table:mie_sys}.

\subsection{Beam Position Monitors}

\begin{figure}[htpb]
\centering
\includegraphics[width=8.5cm]{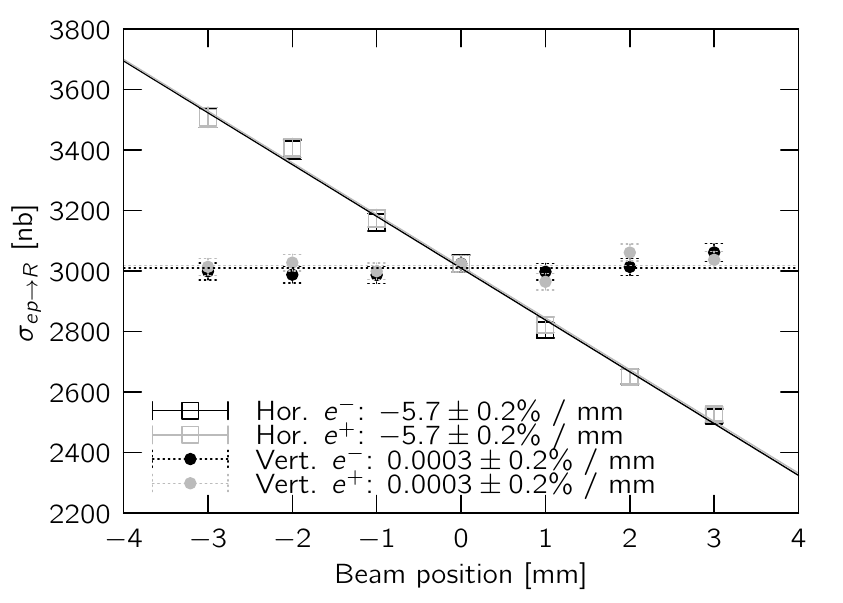}
\caption{\label{fig:symb_bpm_pos} Simulation indicates that the cross-section $\sigma_{ep\rightarrow R}$ is unaffected by changes in the beam's
vertical position, but changes in the horizontal position have a 5.7\%/mm effect, which is independent of beam species.}
\end{figure}

\begin{figure}[htpb]
\centering
\includegraphics[width=8.5cm]{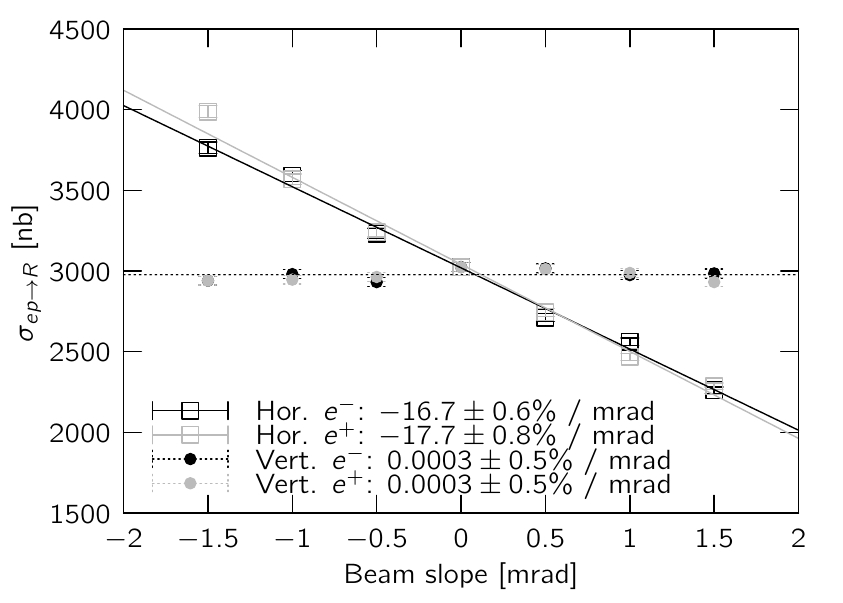}
\caption{\label{fig:symb_bpm_slo} Simulation indicates that the cross-section $\sigma_{ep\rightarrow R}$ is unaffected by changes in the beam's
vertical slope, but changes in the horizontal slope have a 17\%/mrad effect, which is nearly independent of beam species.}
\end{figure}

The largest source of systematic uncertainty came from uncertainty in the beam position monitors (BPMs).
OLYMPUS used two BPMs, one upstream and one downstream of the scattering chamber to locate the beam position and direction
as it passed through the target. This information was used as input to the simulation, so uncertainty
from the BPM measurements could propagate through the simulation to $\sigma_{ep\rightarrow R}$. As seen in
Figs.~\ref{fig:symb_bpm_pos} and \ref{fig:symb_bpm_slo}, $\sigma_{ep\rightarrow R}$ was relatively insensitive
to changes in the beam's vertical position and slope, but changes to the beam's horizontal position and 
slope could produce large effects on the cross-section. However, to produce an error in the relative luminosity
determination, an inaccuracy in the beam position would need to be charge-asymmetric. There are two ways in which 
this could occur:

\begin{enumerate}
\item Inaccuracy of the BPM survey, coupled with different average positions for the $e^+$ and $e^-$ beams,
\item Charge-dependent inaccuracy in the BPMs.
\end{enumerate}

In the case of the former, the accuracy of the BPM measurements was only as good as our knowledge of the
position and orientation of the BPMs themselves. The BPMs were measured during the OLYMPUS optical
survey, and while this survey constrained their positions to better than 0.2~mm, the uncertainty of
the orientation of the BPM axes was conservatively estimated to be 0.5~mm of displacement at a radius
of 55~mm, i.e., $\approx 0.52^\circ$. During OLYMPUS running, the $e^+$ beam had an average vertical 
offset of 0.14~mm in position and 0.44~mrad in slope relative to the $e^-$ beam. Due to uncertainty
in the BPM orientation, these could couple to horizontal shifts of 1.3~\textmu{}m and 1.9~\textmu{}rad, and introduce
uncertainties (according to the simulation results in figures \ref{fig:symb_bpm_pos} and \ref{fig:symb_bpm_slo}) of
0.007\% and 0.034\%.

To assess the charge-dependent accuracy of the BPMs, we used the residuals of fits to charge-independent
calibration data to set an upper bound for any charge-asymmetric effects. We made the conservative assumption
that the non-linearity of these fits, when taken over the full range of possible beam positions, could
contribute 100\% to the charge-asymmetric uncertainty. Therefore, we estimate that the individual BPMs could 
have a charge-asymmetric inaccuracy of no greater than 10~\textmu{}m, and that the readout system could 
produce a fully correlated charge-asymmetric inaccuracy
of no greater than 20~\textmu{}m. The former case could produce a position offset of 7~\textmu{}m (0.04\% uncertainty) 
and a slope  offset of 10~\textmu{}rad (0.17\% uncertainty). The latter case could produce a 20~\textmu{}m horizontal 
position offset introducing an uncertainty of 0.11\%. Combining all of the associated uncertainties in quadrature, 
we find that the BPM uncertainty on the species-relative luminosity is 0.21\%.

\subsection{Magnetic Field at the Target}

The beam position and slope at the target were inferred by assuming a straight-line trajectory of
the beam between the BPMs. However, residual magnetic fields in the target region (described in 
section 6.2 of \cite{Bernauer:2016hpu}) must have produced a slight but opposite curvature for the
trajectories of the $e^+$ and $e^-$ beams, which was unaccounted for. To estimate the size of this
effect, we numerically integrated the equations of motion between the BPM positions interpolating
between measurements of the magnetic field. We found that the target-region magnetic field would
introduce a 3~\textmu{}m offset and a 12~mrad deflection between the $e^-$ and $e^+$ beam, corresponding
to an uncertainty of 0.20\%. 

\subsection{Collimator Geometry}

The simulation made assumptions about the geometry of the SYMB detectors, most crucially the position
and orientation of the collimator aperture for the right calorimeter, which defines the acceptance 
when simulating $\sigma_{ep \rightarrow R}$. The collimator placement in simulation was informed by the
results of the OLYMPUS optical survey. Since the residual magnetic field in the region of the downstream 
beam pipe bends electrons and positrons in opposite directions, geometry error can produce error in the
species-relative luminosity. 

We estimated this error by simulating $\sigma_{ep \rightarrow R}$ for many different positions and orientations
of the right collimator. We found that the dependence on the horizontal position was 0.13\%/mm and 
was smaller than 0.10\%/mm on the vertical position. Moving the collimator forwards and backwards had
a negligible effect on $\sigma_{ep \rightarrow R}$. We gauged the accuracy of the collimator position
and orientation using residuals of fits to the optical survey data, conservatively estimating the position
to be accurate to within 0.5~mm, producing uncertainties of 0.07\% and 0.05\% respectively. We found that 
a rotation about the horizontal axis (moving the face of the collimator up or down) produced a change
of 0.27\%/deg., while a rotation about the vertical axis (moving the face of the collimator left or right)
produced a change of 0.40\%/deg. We conservatively estimated that the collimator orientations were 
accurate to within 0.2$^\circ$, producing uncertainties of 0.05\% and 0.08\% respectively. Adding these
uncertainties in quadrature gives a total of 0.13\% on the species relative luminosity.

\subsection{Event Selection}

\begin{figure}[htpb]
\centering
\includegraphics{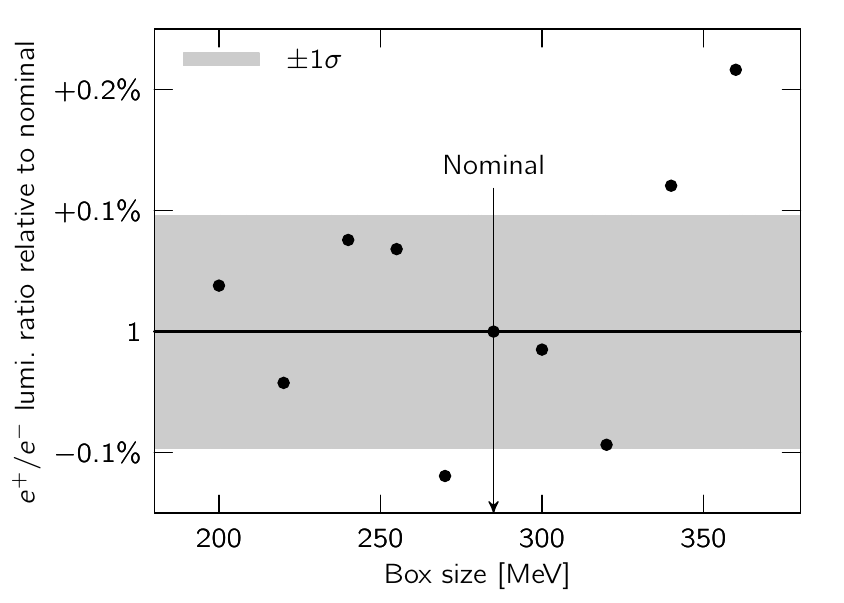}
\caption{\label{fig:boxcut} Adjusting the size of the box-cut region produces
a change in the extracted luminosity ratio of only $\pm 0.10\%$.}
\end{figure}

The total number of counts within the various SYMB signal peaks was estimated by integrating over a 
box-shaped region of the spectrum shown in Fig.~\ref{fig:lm_hist}. Boxes were used for simplicity
after it was shown that shape of the integration region had negligible effect on the total number of
counts, so long as the size of the region was large enough. The signal peaks are well isolated and there
is a negligible contribution from background. We used the same size boxes for both the $(1,1)$ and $(1,3)$
peaks. 

To evaluate the effect of our event selection on the extracted luminosity ratio, the ratio was calculated
for many different sizes of box cuts, both larger and smaller than the nominal size. The results are shown
in Fig.~\ref{fig:boxcut}. The standard deviation of the extracted luminosity ratios was 0.10\%, which we 
quote for the systematic uncertainty due to event selection.

\subsection{Downstream Magnetic Field}

Though the OLYMPUS toroid magnet was designed to keep the magnetic field in the beamline region to
a minimum, there was some residual stray field, which caused the trajectories of electrons
and positrons to bend before entering the SYMBs. This bending is accounted for in the simulation,
but uncertainty in the accuracy of the simulated field model propagates through the simulation to 
uncertainty in $\sigma_{ep\rightarrow R}$. The effect of the magnetic field was
studied in simulation and found to produce a 0.53\% change in the extracted luminosity ratio relative
to a zero-field simulation. We conservatively estimate that our simulation models the effects of this
field correctly to within 10\%, and we quote a systematic uncertainty on the luminosity ratio of 0.05\%. 

\subsection{Higher-Order Corrections}

In the derivation of eq.~\ref{eq:mie_result}, we have neglected terms of higher order than
$\sigma^2$. Based on Fig.~\ref{fig:mie_terms}, this approximation is justified, but we conservatively
quote an uncertainty of 0.05\% for the effect of these neglected terms.

\subsection{Radiative Corrections}

The simulation of $\sigma_{ep\rightarrow R}^\text{sim.}$ used the OLYMPUS radiative event generator,
which calculates a radiative $ep$ cross-section under several different model assumptions. The luminosity
was extracted using each of the various assumptions to test if there was any dependence on model.
The only model choice that was found to affect the luminosity extraction was whether the corrections model 
was based on a $(1+\delta)$ correction or on an exponentiated correction ($\exp(\delta)$). The full scale for this 
difference was 0.05\% on the ratio, i.e., $\sigma_{e^+p\rightarrow R}^\text{sim.} / \sigma_{e^-p\rightarrow R}^\text{sim.}$ 
is 0.05\% higher when exponentiating. We quote a systematic uncertainty of half that difference rounded
up, i.e., 0.03\%.

\subsection{Beam Energy}
\label{ssec:beam_energy}

Uncertainty in the relative beam energy between electrons and positrons produces uncertainty in
the extracted luminosity, via uncertainty in $\sigma_{ep\rightarrow R}^\text{sim.}$. The DORIS
beam energy was monitored over time by using a reference magnet, and was calibrated
through a measurement of the depolarization resonance of the Sokolov-Ternov polarization of the
DORIS beam (see \cite{Steier:2000ui} and \cite{Leemann:2002br} for descriptions of the technique). 
We estimate that the uncertainty in the relative beam energy between electrons and positrons is
500~keV, i.e., $\approx 0.025\%$. To estimate the effect on the luminosity ratio, we evaluated
the change in the Rosenbluth cross-section for scattering at $1.27^\circ$ for 500~keV shifts relative
to a 2.01~GeV beam:
\begin{equation}
  \delta = 1 - \frac{\sigma(E - \delta E,\theta_\text{SYMB})}{\sigma(E + \delta E,\theta_\text{SYMB})}
  \approx 0.10\%,
\end{equation}
where $E=2.01$~GeV, $\delta E=500$~keV, and $\theta_\text{SYMB}=1.27^\circ$.

However, when the luminosity determination is used to normalize a measurement of the $e^+p/e^-p$ cross-section
ratio made with the same beam (as in OLYMPUS), any deviations in the true beam energy affect the luminosity
and the cross section ratio in the same direction, i.e., there is a partial cancellation of the effect.
Rather than adding a $0.10\%$ normalization error over the entire acceptance, the effect of beam
energy uncertainty on the $e^+p/e^-p$ cross-section ratio, i.e., the main OLYMPUS result, is better described with:
\begin{equation}
\delta(\theta) = 1- \frac{\sigma(E - \delta E,\theta)}{\sigma(E + \delta E,\theta)}
\times \frac{\sigma(E + \delta E,\theta_\text{SYMB})}{\sigma(E - \delta E,\theta_\text{SYMB})}
\end{equation}
This uncertainty ranges from 0.04--0.13\% over the OLYMPUS acceptance, and is quoted as the beam energy
uncertainty for the OLYMPUS result \cite{Henderson:2016dea}.

Since beam energy uncertainty and luminosity uncertainty are treated as separate in the OLYMPUS result, 
we make a similar distinction in this work. For convenience, in Table \ref{table:mie_sys}, we present uncertainty 
totals with and without the beam energy uncertainty.

\section{Advantages of the MIE Method}

The chief advantage of the MIE method is that its luminosity extraction comes from
a ratio of two count rates, rather than a single count rate. For a systematic effect
to bias the result, it must affect both count rates differently. This guards against
many forms of detector inefficiency or data acquisition failures, and reduces the degree to which
the extraction is vulnerable to errors in beam position, beam energy, and alignment.

\begin{figure}[htpb]
\centering
\includegraphics{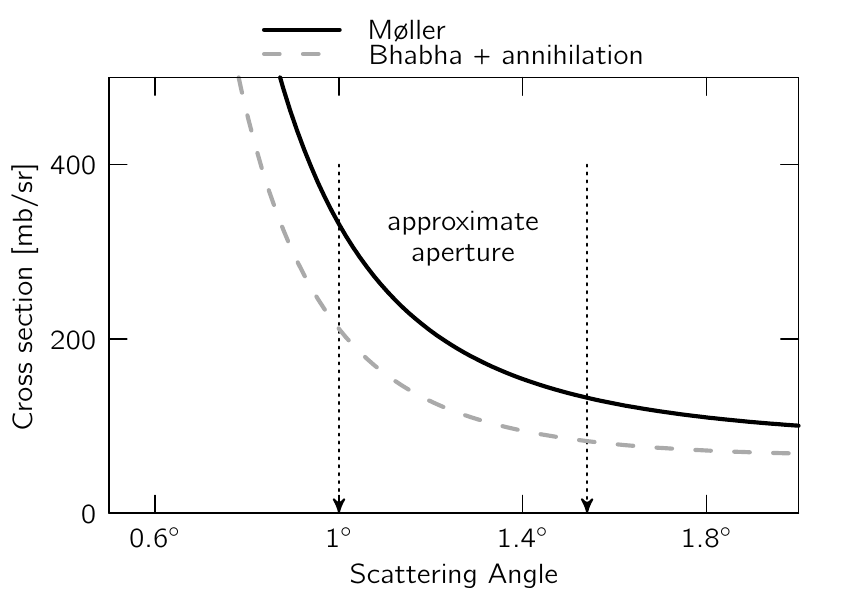}
\caption[Comparison of the M\o ller and Bhabha cross-sections]
{\label{fig:mb_compare} The cross-section for events in the SYMB calorimeters
changes by almost a factor of two between electron running (black solid curve) and 
positron running (gray dashed curve) and has a different angular dependence. This is 
problematic when making a relative luminosity measurement.}
\end{figure}

A second advantage of the MIE method, which is specific to determining relative luminosity
between electron and positron modes, is that the method compares numbers which are of
similar size and of similar dependence on detector acceptance. This is not the 
case when comparing simple rates of M\o ller and Bhabha scattering. 
Fig.~\ref{fig:mb_compare} shows the tree-level cross-sections for M\o ller scattering 
compared with Bhabha scattering (including a small contribution from pair annihilation)
for a 2.01~GeV beam energy in the range of lab-frame scattering angles relevant for OLYMPUS. 
The M\o ller and Bhahba cross-sections are different by more than 60\% within the SYMB acceptance.
Furthermore, M\o ller scattering, being $t$- and $u$-channel, has a different angular dependence
from Bhabha scattering, being $t$- and $s$-channel, so that not only are the relative rates
quite different, this difference depends on the exact detector acceptance. 
It is difficult to control systematic effects when the sensitivity to the detector acceptance
is so large. By contrast, in the MIE method, the large differences between the M\o ller and Bhabha 
cross-sections cancel, while the $ep$ process has a nearly identical cross-section for 
electrons and positrons. This makes the MIE method more robust.

A third advantage is that the MIE method does not rely strongly on simulations of the M\o ller
or Bhabha acceptance. These acceptances cancel between $N_{(1,3)}$ and $N_{(1,1)}$, and so there is
no opportunity for simulation errors to affect the luminosity determination via these terms. 

Future electron- (and/or positron-) scattering experiments using calorimetry to monitor luminosity
between running modes can take advantage of this MIE technique as a means to guard against or
reduce systematic effects. The method employed in OLYMPUS made use of the (1,1) and (1,3) signal peaks 
that were available within the dynamic range of the SYMB ADCs. However, a luminosity monitor designed with the MIE
method in mind should be built to cover many different multi-interaction peaks---(2,2), (3,1), (2,0), etc.---for 
cross checks and to reduce uncertainties. 

\section{Summary}

In this paper, we have presented a method for extracting the relative luminosity between two running
modes using multi-interaction events. This method was used to determine the relative
luminosity between electron and positron datasets in the OLYMPUS experiment \cite{Henderson:2016dea}
and achieved better accuracy than the method of comparing M\o ller and Bhabha rates alone. 
For the specific case of OLYMPUS, we estimate that the method is accurate to within 0.36\% for
a determination of the relative luminosities. The chief advantage of this method, comparing ratios of 
rates rather than rates directly, can easily generalize to other calorimetric luminosity monitors for 
future electron or positron scattering experiments.

\section{Acknowledgements}

We gratefully acknowledge the other members of the OLYMPUS collaboration, who guided the development 
of this work. Discussions with Douglas Hasell, Michael Kohl, Robert Redwine, Uwe Schneekloth, Frank Maas, Dmitry Khaneft,
Rebecca Russell, Brian Henderson, Lauren Ice, and J\"{u}rgen Diefenbach were greatly appreciated.
We wish to thank Frank Brinker and Uwe Schneekloth at DESY, who made possible the DORIS beam
energy calibration. 

This work was supported by the Office of Nuclear Physics of the U.S. Department of Energy, grant No. DE-FG02-94ER40818.

\bibliographystyle{model1-num-names}
\bibliography{references.bib}

\begin{thebibliography}{7}
\expandafter\ifx\csname natexlab\endcsname\relax\def\natexlab#1{#1}\fi
\providecommand{\bibinfo}[2]{#2}
\ifx\xfnm\relax \def\xfnm[#1]{\unskip,\space#1}\fi
\bibitem[{Milner et~al.(2014)Milner, Hasell, Kohl, Schneekloth, Akopov
  et~al.}]{Milner:2013daa}
\bibinfo{author}{R.~Milner}, \bibinfo{author}{D.~Hasell},
  \bibinfo{author}{M.~Kohl}, \bibinfo{author}{U.~Schneekloth},
  \bibinfo{author}{N.~Akopov}, et~al.,
\newblock \bibinfo{title}{{The OLYMPUS Experiment}},
\newblock \bibinfo{journal}{Nuclear Instruments and Methods in Physics Research
  Section A: Accelerators, Spectrometers, Detectors and Associated Equipment}
  \bibinfo{volume}{741} (\bibinfo{year}{2014}) \bibinfo{pages}{1--17}.
\bibitem[{Henderson et~al.(2016)}]{Henderson:2016dea}
\bibinfo{author}{B.~Henderson}, et~al.,
\newblock \bibinfo{title}{{Hard Two-Photon Contribution to Elastic
  Lepton-Proton Scattering: Determined by the OLYMPUS Experiment}},
\newblock \bibinfo{journal}{Phys. Rev. Lett.} \bibinfo{volume}{accepted}
  (\bibinfo{year}{2016}).
\bibitem[{P{\'e}rez~Benito et~al.(2016)P{\'e}rez~Benito, Khaneft, O'Connor,
  Capozza, Diefenbach, Gl{\"a}ser, Ma, Maas, and Pi{\~n}eiro}]{Benito:2016cmp}
\bibinfo{author}{R.~P{\'e}rez~Benito}, \bibinfo{author}{D.~Khaneft},
  \bibinfo{author}{C.~O'Connor}, \bibinfo{author}{L.~Capozza},
  \bibinfo{author}{J.~Diefenbach}, \bibinfo{author}{B.~Gl{\"a}ser},
  \bibinfo{author}{Y.~Ma}, \bibinfo{author}{F.~Maas}, \bibinfo{author}{D.~R.
  Pi{\~n}eiro},
\newblock \bibinfo{title}{{Design and Performance of a Lead Fluoride Detector
  as a Luminosity Monitor}},
\newblock \bibinfo{journal}{Nucl. Instrum. Meth.} \bibinfo{volume}{A826}
  (\bibinfo{year}{2016}) \bibinfo{pages}{6--14}.
\bibitem[{Benisch et~al.(2001)Benisch, Bernreuther, Devitsin, Kozlov, Potashov,
  Rith, Terkulov, and Weiskopf}]{Benisch:2001rr}
\bibinfo{author}{T.~Benisch}, \bibinfo{author}{S.~Bernreuther},
  \bibinfo{author}{E.~Devitsin}, \bibinfo{author}{V.~Kozlov},
  \bibinfo{author}{S.~Potashov}, \bibinfo{author}{K.~Rith},
  \bibinfo{author}{A.~Terkulov}, \bibinfo{author}{C.~Weiskopf},
\newblock \bibinfo{title}{{The luminosity monitor of the HERMES experiment at
  DESY}},
\newblock \bibinfo{journal}{Nucl. Instrum. Meth.} \bibinfo{volume}{A471}
  (\bibinfo{year}{2001}) \bibinfo{pages}{314--324}.
\bibitem[{Bernauer et~al.(2016)}]{Bernauer:2016hpu}
\bibinfo{author}{J.~C. Bernauer}, et~al.,
\newblock \bibinfo{title}{{Measurement and tricubic interpolation of the
  magnetic field for the OLYMPUS experiment}},
\newblock \bibinfo{journal}{Nucl. Instrum. Meth.} \bibinfo{volume}{A823}
  (\bibinfo{year}{2016}) \bibinfo{pages}{9--14}.
\bibitem[{Steier et~al.(2000)Steier, Byrd, and Kuske}]{Steier:2000ui}
\bibinfo{author}{C.~Steier}, \bibinfo{author}{J.~Byrd},
  \bibinfo{author}{P.~Kuske},
\newblock \bibinfo{title}{{Energy calibration of the electron beam of the ALS
  using resonant depolarization}},
\newblock in: \bibinfo{booktitle}{{Particle accelerator. Proceedings, 7th
  European Conference, EPAC 2000, Vienna, Austria, June 26-30, 2000. Vol.
  1-3}}, pp. \bibinfo{pages}{1566--1568}.
\bibitem[{Leemann et~al.(2002)Leemann, Boege, Dehler, Schlott, and
  Streun}]{Leemann:2002br}
\bibinfo{author}{S.~C. Leemann}, \bibinfo{author}{M.~Boege},
  \bibinfo{author}{M.~Dehler}, \bibinfo{author}{V.~Schlott},
  \bibinfo{author}{A.~Streun},
\newblock \bibinfo{title}{{Precise beam energy calibration at the SLS storage
  ring}},
\newblock in: \bibinfo{booktitle}{{Particle accelerator. Proceedings, 8th
  European Conference, EPAC 2002, Paris, France, June 3-7, 2002}}, pp.
  \bibinfo{pages}{662--664}.

\end{thebibliography}

\end{document}